\documentclass[twocolumn,showpacs,preprintnumbers,amsmath,amssymb,prl]{revtex4}
\usepackage{graphicx}% Include figure files
\usepackage{bm}% bold math
\usepackage{natbib}
\begin{document}

\title{Two-band superconductivity of bulk and surface states in Ag thin films on Nb}
\author{Tihomir Tomanic$^{1}$}
\author{Michael Schackert$^{1}$}
\author{Wulf Wulfhekel$^{1}$}
\author{Christoph S\"{u}rgers$^{1}$}
\email{christoph.suergers@kit.edu}
\author{Hilbert v. L\"ohneysen$^{1,2}$}
\affiliation{$^1$Physikalisches Institut, Karlsruhe Institute of Technology, P.O. Box 6980, 76049 Karlsruhe, Germany}
\affiliation{$^2$Institut f\"ur Festk\"orperphysik, Karlsruhe Institute of Technology, P.O. Box 3640, 76021 Karlsruhe, Germany}

\begin{abstract}
We use epitaxial strain to spatially tune the bottom of the surface-state band $E_{\rm SS}$ of Ag(111) islands on Nb(110). Bulk and surface-state contributions to the Ag(111) local density of states (LDOS) can be separated with scanning tunneling spectroscopy. For thick islands ($\approx$\, 20 nm), the Ag surface states are decoupled from the Ag bulk states and the superconductive gap induced by proximity to Nb is due to bulk states only. However, for thin islands (3-4 nm), surface-state electrons develop superconducting correlations as identified by a complete energy gap in the LDOS when $E_{\rm SS}$ is smaller than but close to the Fermi level. The induced superconductivity in this case is of two-band nature and appears to occur when the surface-state wave function reaches down to the Ag/Nb interface.
\end{abstract}
%\date{\today}

\maketitle
With the discovery of topological insulators \cite{kane_quantum_2005,konig_quantum_2007,hasan_textitcolloquium_2010}, topologically protected surface states have become one of the most challenging topics in today's condensed-matter physics. In bulk insulating materials, a nontrivial band inversion leads to conducting surface states. A conventional s-wave superconductor in proximity to a topological insulator \cite{wang_interplay_2012} might  carry exotic superconducting states with non-Abelian Majorana states \cite{fu_superconducting_2008}. It has been proposed that surface-state electrons are strongly affected by the inversion asymmetry of the surface, generating large Rashba  couplings \cite{potter_topological_2012}. Proximity-induced superconductivity has been studied for heavy metals like Ag, Au, and Pb  \cite{le_sueur_phase_2008,wang_proximity-induced_2009,wolz_evidence_2011,serrier-garcia_scanning_2013} and electrostatic control of superconductivity (SC) has been reported for proximity-induced SC generated by metallic tin nanoparticles on graphene \cite{allain_electrical_2012} and for electron accumulation at the surface of semiconducting transition-metal dichalcogenides \cite{ye_superconducting_2012,jo_electrostatically_2015}. Recently, magnetoresistance measurements on nanoplates of the In-doped topological superconductor candidate SnTe revealed the role of surface states despite the high bulk carrier density \cite{shen_revealing_2015}. Here we report on the interplay between bulk and surface states of Ag with regard to SC induced by proximity to superconducting Nb. We apply epitaxial strain to tune the energy of the surface-state band, allowing to disentangle the relative roles of bulk and surface states in superconducting Ag.

The role of surface states of a bulk superconductor that do not hybridize with bulk states has become a new twist with the discovery of topologically protected surface states. Ginzburg  already considered SC of electrons in two-dimensional (2D) surface states \cite{Ginzburg_Kirzhnits_1964}. However, for simple conventional SC, this situation is hard to find in nature. The range of materials can be considerably expanded by considering proximity-induced SC. Thus, we have investigated (111)-oriented Ag nano-islands on a superconducting Nb (110) single crystal ($T_c$ = 9.1 K) to address this question. The Ag(111)/Nb(110) system fulfills two prerequisites: First of all, Ag islands of less than $\approx 100$\,nm thickness become fully penetrated by the superconductive order parameter \cite{stalzer_field-screening_2006}. Secondly, the Ag(111) surface exhibits an intrinsic Shockley-type electronic surface state (SS) in the projected \textit{sp} band gap of the Ag band structure \cite{kevan_high-resolution_1987}. The bottom of the SS band appears at an energy $E_{\rm SS} \approx -65$\, meV below the Fermi level $E_{\rm F}$ \cite{reinert_direct_2001,kliewer_dimensionality_2000,limot_surface-state_2003,limot_surface-state_2005}. SS electrons on Ag(111) have a parabolic dipersion with an effective mass $m^{\ast} = 0.42\,m_e$ ($m_e$: electron mass) \cite{becker_theoretical_2006}. Their wave function decays exponentially into the bulk with a decay length $\beta^{-1}$ = 1.8 - 2.8 nm \cite{hsieh_probing_1985,bendounan_modification_2003}.

Single-crystalline (111)-oriented Ag islands deposited on Nb(110) at ambient or elevated substrate temperatures $T_S$ experience considerable thermal strain when cooled to low temperatures due to the different thermal expansion coefficients of Ag and Nb as reported earlier \cite{tomanic_local-strain_2012}. The tensile strain in the Ag film plane strongly shifts the bottom SS band edge across $E_{\rm F}$ thereby depopulating the SS \cite{neuhold_depopulation_1997,sawa_thickness_2009,jiang_scanning_2001,tomanic_local-strain_2012}. Hence, this effect allows to separate the contribution of electrons from bulk and surface states to the LDOS at $E_{\rm F}$ and to investigate the effect of bulk SC on the normal SS. 
   
The experiments have been performed in ultra-high vacuum (base pressure $p < 10^{-8}$\,Pa). Ag(111) islands deposited at $T_S$ = 600 K grow on a one-monolayer thick (0.236 nm) Ag wetting layer on Nb(110) \cite{ruckman_growth_1988}. The $\approx 2^{\circ}$ miscut of the Nb(110) substrate yields Ag islands of wedge-like shape with  different size and height distributions and hence locally varying $E_{\rm SS}$ \cite{tomanic_local-strain_2012}. Scanning tunneling microscopy (STM) and scanning tunneling spectroscopy (STS) were performed in-situ at $T = 5$\,K as well as in a Joule-Thomson STM \cite{zhang_compact_2011} at $T = 0.7-0.9$\, K using in-situ cleaned tungsten tips with the bias voltage $V$ applied to the sample. $dI/dV$ spectra representing the single-particle LDOS of the samples were acquired using a lock-in amplifier with a modulation voltage $V_{mod}$ at a frequency of 3.21 kHz.  
\begin{figure}
\includegraphics[width=1\columnwidth,clip=]{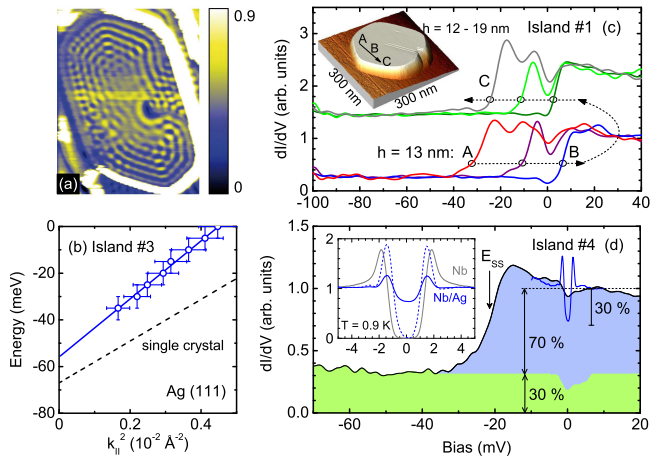}
\caption[]{(a) $dI/dV$ map of Ag island \#3 ($92 \times 120\, {\rm nm}^2$, height 18-20 nm) at $T$ = 5 K and $-14$\, mV. (b) Energy dispersion of 2D surface electrons determined from the $dI/dV$ map. Dashed line indicates data of Ag single crystals \cite{becker_theoretical_2006}.  (c) $dI/dV$ spectra along the path A $\rightarrow$ B  $\rightarrow$ C at $T$ = 5 K. Data are vertically shifted for clarity. (d) $dI/dV$ in the superconducting state at 0.9 K obtained with low (black curve) and high (blue curve) resolution. Colored areas indicate contributions from bulk (green) and surface states (blue) for the low-resolution data. Inset shows the high-resolution data (blue curve) and $dI/dV$ after subtracting the value at $V = 0$ and subsequent rescaling (dashed blue curve) for comparison with Nb (gray).}
\label{fig1}
\end{figure}

The map of $dI/dV$ values [Fig. \ref{fig1}(a)] shows the characteristic variation of the LDOS due to scattering of 2D SS electrons at defects and at the island edges and interference of the incident and back-scattered wave \cite{crommie_imaging_1993,li_local_1997,burgi_two-dimensional_2002,sawa_thickness_2009}. The observation of the interference pattern indicates a lateral electron mean-free path which is at least as long as the size of the island (120 nm). From $dI/dV$ maps we obtain the energy dispersion of SS electrons [Fig. \ref{fig1}(b)] and $m^{\ast} = 0.3\, m_e$ for this particular island. $m^{\ast}$ being smaller than $0.42\, m_e$ of Ag(111) single crystals has been also reported for strained Ag films on Si(111) $7 \times 7$ \cite{sawa_thickness_2009}. The determination of $E(k_{\parallel})$ is complicated due to the variation of  $E_{\rm SS}$ across the island and the details of the hybridization between Ag bulk states and the Nb substrate. Fig. \ref{fig1}(c) shows $dI/dV$ spectra over a large energy range taken at different positions on top of island \#1. The step-like behavior represents the bottom SS band edge of the 2D LDOS \cite{kliewer_dimensionality_2000,limot_surface-state_2003,limot_surface-state_2005}. The step shifts toward $V = 0$, representing $E_{\rm F}$, when the tip is moved from the island edge to the center (A $\rightarrow$ B) and back to negative voltages (B $\rightarrow$ C) due to the laterally inhomogeneous thermal strain as discussed in detail in an earlier publication \cite{tomanic_local-strain_2012}. We find regions where the band edge is shifted to above $E_{\rm F}$ ($V > 0$) (blue curve). It is important to note that along the path A-B-C the height of the island stays at $h = 13$\,nm, corroborating our previous conclusion that it is the strain that is responsible for the shift in $E_{\rm SS}$. Of course, we cannot exclude a minor contribution of defects and impurities to the local strain. 

Fig. \ref{fig1}(d) shows two spectra acquired with different voltage resolutions at $T = 0.9$\,K. Over a large range of bias voltage and with a modulation voltage $V_{mod} =$ 1 mV$_{\rm rms}$ (black curve), the SS band edge is clearly resolved at -20 mV bias. Both, SS and bulk states, contribute to the differential conductance according to the Tersoff-Hamann model \cite{tersoff_theory_1983}. The relative contributions to the total LDOS arising from SS ($\approx$ 70 \%, blue area) and bulk states ($\approx$ 30 \%, green area) are in agreement with earlier investigations on Ag(111) single crystals \cite{kliewer_dimensionality_2000,burgi_two-dimensional_2002,limot_surface-state_2003}. The blue curve obtained with higher resolution $V_{mod} =$ 0.1 mV$_{\rm rms}$, shows the characteristic minimum at $E_{\rm F}$ ($V$ = 0) and two peaks at $V \approx \pm \Delta_0/e$, where $2\Delta_0$ is the superconductive energy gap, due to the proximity-induced SC of Ag. When the SS contribution is subtracted from the data and rescaled to obtain the same $dI/dV$ value at $V \gg \Delta_0/e$ as on Nb we find a slightly reduced gap of 1.2 meV compared to $\Delta_0 = $\, 1.5 meV observed on the Nb surface, see inset Fig. \ref{fig1}(d). All islands are thinner than the coherence lengths in Ag, estimated to 1600 nm or 250 nm in the clean or dirty limit at $T$\, = 1 K \cite{stalzer_field-screening_2006}, respectively, and become fully superconducting when $E_{\rm SS} \gg E_{\rm F}$. We note that in proximity-coupled Nb(110)/Ag(111) double layers the break-down field for the Meissner effect can be described by the clean-limit expression \cite{stalzer_field-screening_2006}. 

\begin{figure}
\includegraphics[width=0.9\columnwidth,clip=]{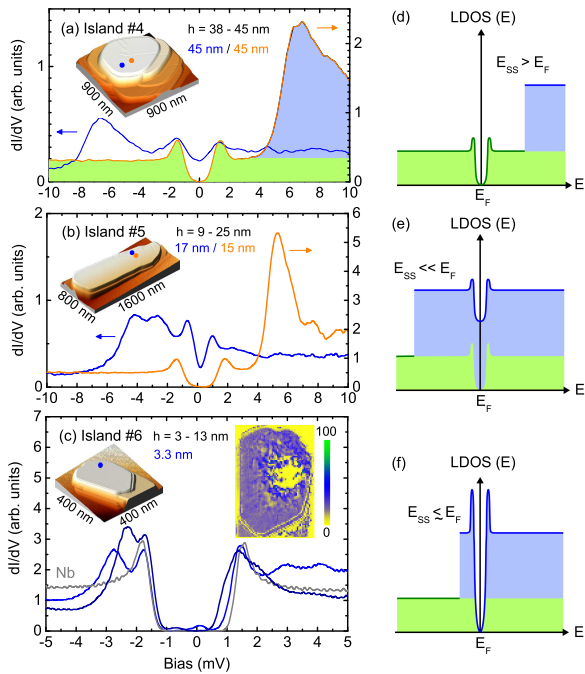}
\caption[]{(a-c) $dI/dV$ tunneling spectra of Ag islands on Nb(110) at $T$ = 0.7 - 0.9 K. Colored points in the 3D STM images indicate measuring positions. (c) Gray curve shows a spectrum obtained on the Nb surface, blue curves show two spectra with $E_{\rm SS} < E_{\rm F}$. Right inset shows a $dI/dV$ map on a relative scale 0-100 \% at $V$ = -1.0 mV. (d-f) Schematic of the LDOS for different $E_{\rm SS}$.} 
\label{fig2}
\end{figure}

In the following, we exemplarily discuss $dI/dV$ spectra obtained with high resolution on three islands of different size. Fig. \ref{fig2}(a) shows the usual behavior observed for island \#4. If the SS band is completely unoccupied with a band edge at $V \approx$ 5 mV well above $E_{\rm F}$ (orange curve), the superconductive gap around $V = 0$ is fully developed in the LDOS of Ag bulk states, schematically shown in Fig. \ref{fig2}(d). We define a ``gap depth'' 
\begin{equation}
GD = 1- \left|\frac{dI/dV_{V=0}}{dI/dV_{\left|V_0\right|}}\right| 
\end{equation}
(for small voltages $\left|V_0\right|$ just above $\Delta_0/e$) which in this case is 100 \%. Vice versa, if the SS band edge is at $V \approx -8$\, mV (blue curve), $GD \approx$\, 30 \% perfectly agrees with the contribution from the bulk band to the total LDOS, cf. Figs. \ref{fig1}(d) and \ref{fig2}(e). Hence, electrons in Ag bulk states become superconducting by the proximity effect with Nb but the LDOS of SS is not affected.

A similar behavior is observed for island \#5 with a lower height of 9 - 25 nm shown in Fig. \ref{fig2}(b). Here, the SS band edge evolves into a striking maximum. This behavior is more pronounced when the islands are thin and the SS band edge is close to the Fermi level, see below. Again, the unoccupied SS band above 5 mV (orange curve) is unaffected by superconductivity and $GD$ = 100 \% because close to $E_{\rm F}$ only bulk states are occupied. However, we also find locations on this island where the SS band is occupied and has a band edge at $\approx -5$\, mV (blue curve) but shows a deeper minimum at $V = 0$ and a corresponding $GD$ of 50 \%. This suggests that in this case part of the SS electrons of Ag(111) are affected by SC and contribute to the superconducting condensate. 

The gap in the SS band is even more pronounced when the island thickness is further reduced. Fig. \ref{fig2}(c) shows two spectra of island \#6 where the SS band edge is close to -3 mV and is occupied. In this case, the superconductive gap is fully developed, i.e., $GD$ = 100 \%. The appearance of a gap in the LDOS of both electronic bulk \textit{and} surface states - schematically shown in Fig. \ref{fig2}(f) - immediately suggests that the SS electrons fully develop superconductive correlations. For this island we do not find regions where the SS band edge is shifted to above $E_{\rm F}$ presumably due to the small island thickness. For Ag(111) films on Cu(111) and Au(111), the SS considerably shifts to \textit{lower} energies when the thickness becomes comparable to $\beta ^{-1}$ \cite{bendounan_modification_2003,didiot_imaging_2005}. The $dI/dV$ map [Fig. \ref{fig2}(c), inset] was recorded at $V$ = -1.0 mV. The yellow colored areas indicate regions where $dI/dV$ = 0 corresponding to a $GD$ = 100 \% which is only observed in the center of the island and on the surrounding Nb substrate surface. 

\begin{figure}
\includegraphics[width=\columnwidth,clip=]{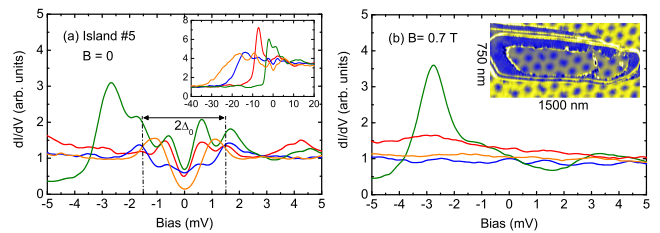}
\caption[]{$dI/dV$ tunneling spectra obtained on Ag island \#5 in (a) zero field and in (b) magnetic field. Inset shows the step-like behavior arising from the lower surface-state band edge evolving into a maximum close to $V = 0$. The STM image in (b) shows the Abrikosov lattice of superconducting vortices.}
\label{fig3}
\end{figure}

Figure \ref{fig3}(a) shows a few more spectra obtained on island \#5 that display additional features. In some cases the gap is deeper than 30 \% of the Ag bulk states although the SS state is well below $E_{\rm F}$, cf. orange curve and Fig. \ref{fig2}(b). In addition, distinct maxima are seen at $\pm 0.6$\, mV within the band gap $2\Delta_0$. These subgap features disappear in a perpendicular magnetic field of $B = 0.7$\, T proving that they are due to SC, see Fig. \ref{fig3}(b). 
%From the vortex diameters and a BCS coherence length $\xi_0 $\, = 40 nm of Nb we estimate Ginzburg-Landau coherence lengths $\xi_{\rm GL}^{\rm Nb} \approx 45 \pm 5$\,nm and $\xi_{\rm GL}^{\rm Ag} \approx 50 \pm 5$\,nm \cite{eskildsen_vortex_2002}, respectively, and an electron mean free path $l \approx 3 \xi_0$ = 120 nm \cite{renner_scanning_1991}. For this value of \textit{l}, a reduced gap (dubbed ``minigap'') can be estimated as 1.26 meV \cite{belzig_local_1996}, in good agreement with our experimental value of 1.2 meV. 

The inset of Fig. \ref{fig3}(a) shows a behavior already mentioned above: The step in $dI/dV$ evolves into a sharp maximum when the SS band edge shifts toward $E_{\rm F}$, see also Fig. \ref{fig2}(b). This cannot be due to the confinement of electrons generating quantum-well states at much higher binding energies for this island thickness \cite{neuhold_depopulation_1997}. Likewise, the maxima cannot be attributed to adsorbate-induced bound states split off from the SS band edge which should appear independent of the value of $E_{\rm SS}$, which is not observed \cite{limot_surface-state_2005}.      

The main observation of Figs. \ref{fig2}(a) to \ref{fig2}(c) suggests that, due to the finite SS decay length $\beta ^{-1}$, electrons in the SS become superconducting only for very thin Ag islands. The 2D SS and three-dimensional bulk states are orthogonal. In principle, interactions between electrons in these states could occur via scattering \cite{potter_topological_2012} at the island rim, by defects, or by the potential barrier at the Ag/Nb interface, provided that the Ag thickness is in the range of $\beta^{-1}$ = 2.8 nm \cite{hsieh_probing_1985}. For island \#6 we estimate an amplitude of 30 \% of the SS wave function at the Ag/Nb interface, compared to the surface. However, the $dI/dV$ map of Fig. \ref{fig2}(c) (inset) shows that $dI/dV$ = 0 is obtained in a confined region in the center of the island, i.e., $GD$ = 100 \%. Hence, scattering at the island rim or at defects does not seem to play a role in inducing SC.

\begin{figure}
\includegraphics[width=0.7\columnwidth,clip=]{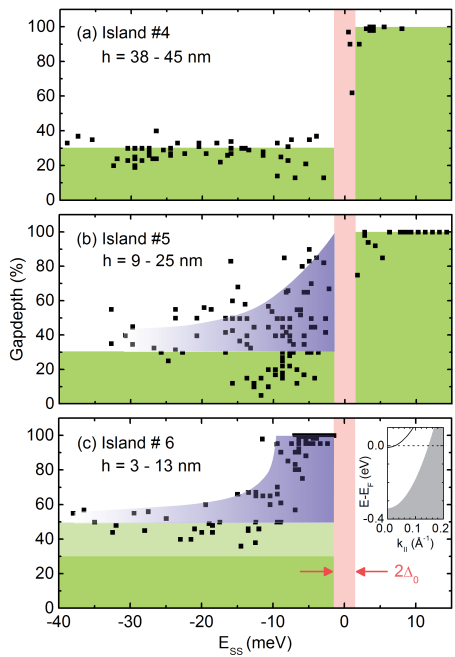}
\caption[]{$GD$ vs. $E_{\rm SS}$ for Ag islands of different thickness. Red: width of the superconductive gap $2\Delta_0 \approx 3\, {\rm meV}$ of Nb. Green and blue areas indicate the contribution of bulk and surfaces states, respectively. Inset in (c) shows the energy dispersion of the SS state band [cf. Fig. \ref{fig1}(b)] for $E_{\rm SS}$ = -10 meV (solid line) and the projected bulk bands (shaded area).}
\label{fig4}
\end{figure}

To corroborate our findings, we have investigated a large number of spectra and have determined the $GD$ in dependence of the SS band edge $E_{\rm SS}$. Figure \ref{fig4} shows that for the thickest island \#4 the $GD$ at $E_{\rm F}$ is about 30 \% for all occupied SS states with $E_{\rm SS} < 0$, and about 100\% for unoccupied SS with $E_{\rm SS} > 0$ as shown above, with a very small transition range of the order of $2 \Delta_0$. For the thinner island \#5, the spectra for $E_{\rm SS} < 0$ show a large scatter of $GD$ with enhanced $GD$ of 30 - 90 \%, with a tendency of larger $GD$ values when $E_{\rm SS}$ approaches $E_{\rm F}$. The most intriguing behavior is observed for island \# 6 where $GD$ of 100 \% is obtained even for $E_{\rm SS}$ in the range between $-$10 meV and $-\Delta_0$, i.e., at energies yet below the superconductive gap. In this case the LDOS shows a fully developed gap in the electronic bulk \textit{and} surface states. Furthermore, we note that for this sample even for large negative values of $E_{\rm SS}$ the $GD$ in bulk states exceeds somewhat the expected 30\%-value indicated by the light-green area in Fig. \ref{fig4}(c). 

Let us summarize the experimental facts: (i) in the normal state, the density of SS is about two times larger than that of Ag bulk states, (ii) for thick islands, surface states do not participate in the proximity-induced SC leading to a $GD$ of 100 \% when SS are not occupied, but only 30 \% when they are, (iii) if the Ag-film thickness is smaller than the SS decay length, these surface states do partake in SC provided that $E_{\rm SS} \leq E_{\rm F}$ is within a window of $\approx$\, 10 meV of $E_{\rm F}$. Spin-orbit (SO) interaction cannot be the reason because the upper limit of the SO splitting is 1.9 meV \cite{nicolay_spin-orbit_2001} and SO interaction cannot lift the orthogonality between Ag(111) surface and bulk states. Since single-particle states do not couple, one has to resort to many-body effects associated with SC, e.g., interband pairing between electronic surface and bulk states, or impurity or defect scattering \cite{potter_topological_2012}. It is important to point out that interband paring of course requires a finite pairing potential. This is different to the usual proximity-induced SC arising from a finite pairing amplitude in the \textit{absence} of a pairing potential in the nonsuperconducting metal in contact with a superconductor.   

The first possibility of interband pairing is that electron-phonon coupling between states of different bands is enhanced, while the Cooper pairs are still formed within a single band  \cite{geerk_observation_2005}. Applying this idea to Ag bulk and surface states, we note that the SS is a state located in a potential well confined by the crystal band gap and the surface barrier on the vacuum side. Oscillations of the electron-phonon coupling strength $\lambda$ as a function of quantum-well thickness $d_Q$ have been observed \cite{luh_large_2002}. Hence, it is concievable that $\lambda(d_Q)$ accidentally has a maximum at the experimental $d_Q$. An enhanced electron-phonon interaction has recently reported for Pb islands on Cu(111) \cite{schackert_local_2015}.

Furthermore, electron-phonon scattering is dominant over electron-electron scattering for SS close to $E_{\rm F}$ \cite{kliewer_dimensionality_2000,becker_theoretical_2006,eiguren_role_2002}. For Ag(111), a Rayleigh surface mode ${\rm S_1}$ gives rise to a low-energy peak in the Eliashberg function at $\approx$\, 8 meV \cite{ponjee_experimental_2003}. It is conceivable that this mode plays an important role in enhancing the stability of SC for $E_{\rm SS} < -\Delta_0$. We also checked the possiblity of a strongly enhanced LDOS supporting SC due to a Rashba splitting of the Ag(111) SS by $\Delta k = 0.005$\, \AA $^{-1}$ \cite{ishida_rashba_2014}. Such an enhancement would occur only for $E-E_{\rm F} < 80 \mu$V independent of $E_{SS}$ and does not explain the observed behavior. A more exotic possiblity that Cooper pairs are formed of two quasiparticles from two different bands appears possible in principle, but to our knowledge no calculations for this scenario exist. 

In conclusion, we have shown that electrons in Ag(111) surface states become superconducting when the surface-state wave function reaches down to the Nb/Ag interface which serves as a potential barrier for scattering, and, additionally, the lower band edge of the surface states falls within a range of 10 meV below $E_{\rm F}$. The enhanced electron-phonon interaction close to the Fermi level might facilitate interband pairing. Although a possible new collective phonon mode is hard to detect in the Ag nano-island, superconductivity acts as a ``smoking gun'' proving the existence of such a mode. The presence of an interband pairing potential points to an electronic interaction between the bulk superconductor Nb and the Ag(111) surface states that differs from the usual proximity effect between bulk Ag and Nb.

We gratefully acknowledge financial support from the Kompetenznetzwerk ``Functional Nanostructures'' of the Baden-W\"urttemberg Stiftung and thank D. Beckmann, W. Belzig, H. Kroha, E. Scheer, J. Schmalian, R. Heid, and M. Wenderoth for several fruitful discussions.

%\bibliography{NbAg_SS_resubm}

\end{document}